\documentclass[12pt,a4paper]{article}
\usepackage{amsmath}
\usepackage{graphicx}
\usepackage{times}
\begin{document}
\begin{titlepage}
\title{Implications of the LHC results for the cosmic data interpretations}
\author{S.M. Troshin,
 N.E. Tyurin\\[1ex]
\small  \it Institute for High Energy Physics, NRC ``Kurchatov Institute"\\
\small  \it Protvino,  142281, Russian Federation}
\normalsize
\date{}
\maketitle

\begin{abstract}
Absence of the signals of the physics beyond the Standard Model at the available LHC energies means
that the several of cosmic data peculiarities could be associated with an emergency  
of the new scattering mode.  This reflective 
scattering mode
can be used for explanations of the measurements performed under the study of
extensive air showers (EAS). \\[2ex]
\end{abstract}
\end{titlepage}
\setcounter{page}{2}

\section*{Introduction}
The  cosmic rays studies are an
 important data source in astrophysics.
They simultaneously provide a window on the future results of accelerator
studies  at the energies already covered by  and  far beyond of  the  LHC \cite{lipari}. 
The primary flux of cosmic particles  is  composed mainly from the protons. 
The measurements are performed  with the extensive air showers developed in the atmosphere, since
beyond the energy $10^{14}$ eV this flux has low intensity. Thus, the cosmic rays studies
suffer from low intensity of the primary  flux at the energies higher than $10^{14}$ eV 
and related necessity to exploit the particular model extrapolations
for the interpretations of the EAS results.  

The number of
model interpretations can be reduced due to the results obtained at the accelerators. 
To be more specific, we consider  a knee in the energy spectrum of cosmic particles.
%\begin{figure}[thb]
%\begin{center}
%\resizebox{8cm}{!}{\includegraphics*{flux}}
%\includegraphics[width=90mm]{flux.eps}
%\end{center}
%\small{\caption{Scaled energy spectrum of the cosmic rays, figure from  \cite{engel}.}}
%\label{ts:fig1}
%\end{figure}
The
energy spectrum  follows a simple power-like law $F(E)=cE^{-\gamma}$ but it changes
the slope in the energy region of $\sqrt{s}=3-6$ TeV\cite{crrev,her} and becomes steeper: 
index $\gamma$ increases from
$2.7$ to $3.1$. Just this change is called a knee. 

The dominant interpretation associates this change with astrophysics.
However, it can also be related to the dynamics of the primary particle interactions
with the atmosphere. 
In particular, the knee  was interpreted as a result of appearance
 of the new particles with a small inelastic cross--section and/or
small inelasticity. These new particles were associated with effects of the
supersymmetry, technicolor and other new physics  beyond the Standard Model (SM) (cf. e.g. \cite{dixi}). 
But, nowadays the latter option seems to be disfavored by the LHC results \cite{dent}.
Physics beyond the SM has not been found so far
in the energy range $\sqrt{s}=3-6$ TeV.  This fact can have the two alternative interpretations. To adopt an astrophysical
origin of the knee and other peculiarities, it should be accepted that  cosmic rays content changes  with energy  
from light to a more heavy particles \cite{dent1}. 
This option seems to be presently a favorable one as it was stressed in the above paper.

 There is a second possibility which is not excluded by the data and, moreover, the LHC data have pointed out its existence \cite{alkin}. This option does not presume change in the primary cosmic rays composition.
It continues to correlate the observed effects 
with dynamics of the primary cosmic ray interactions with atmosphere 
and gradually turning on new scattering mode with the energy. 
Namely, there 
is a possibility to consider the observed effects in EAS as the manifestations
of emerging the reflective scattering mode with the energy increase \cite{inja}.

We discuss this possibility in  the next section.
 \section{The reflective scattering mode and interpretation of the EAS data}

The elastic scattering $ S$-matrix element  can be expressed through the elastic scattering amplitude $f(s,b)$ by the relation $S(s,b)=1+2if(s,b)$\footnote{At high  energies the real part of the scattering amplitude is small and can  be neglected,  allowing the substitution $f\to if$.}.  It can also be represented in the form
\[S(s,b)=\kappa(s,b)\exp[2i\delta(s,b)],\]
which includes   two real functions $\kappa(s,b)$ and $\delta(s,b)$. The function $\kappa$ ($0\leq \kappa \leq 1$) is a transmission coefficient 
\cite{chng}, and
the value $\kappa=0$ corresponds to a complete absorption of the initial state. 

We consider scattering at high energies and use therefore
an approximation of the pure imaginary scattering amplitude when the function $S(s,b)$ is real.
The selection of elastic scattering mode, namely, absorptive  or reflective one, is governed by the phase $\delta(s,b)$. 
The standard assumption is   that $ \kappa(s,b)\to 0$  at the fixed impact parameter $b$ and $s\to \infty$. This is called a black disk limit 
since  in this case the elastic scattering amplitude  $f(s,b)$ has a completely absorptive origin and
is limited by the inequality $f(s,b)  \leq 1/2$.  

There is  another option, namely, the function $ S(s,b)\to -1$ at fixed $b$  and $s\to \infty$, 
i.e. transmission coefficient $\kappa \to 1$ and $\delta = \pi/2$. The mechanism corresponding to such a dependence can be described
 as  a pure reflective scattering \cite{inja}. 
The distinct feature of the reflective scattering mode is that the amplitude varies in the range  $1/2  < f(s,b) \leq 1$, as it is allowed by unitarity \cite{phl1}.
Model
estimates show that
new scattering mode starts to develop right beyond the Tevatron energy range, i.e. at
$\sqrt{s_0}> 2$ TeV \cite{inja}, which corresponds to the energy in the laboratory system
$E_0\simeq 2$ $PeV$. The energy value $s_0$ is determined by the relation $\kappa(s,b=0)=0$.
This reflective scattering mode starts to develop at the LHC and it has been revealed in the  data \cite{alkin} at $\sqrt{s}=7$ TeV.
 The description of this mode can be found  in \cite{inja,phl1,phl}.
The most important feature
is a self-damping of the
inelastic channels  at small values of the impact parameter.  
The reflective scattering leads to the asymptotic limit $P(s,b=0)\to 1$ at $s\to\infty$, where $P$ is
a probability of the absence of the inelastic interactions, namely,
$P(s,b)\equiv \kappa ^2(s,b)=1-4h_{inel}(s,b)$, where the function $h_{inel}(s,b)$ is a contribution of the inelastic channels
to the unitarity relation. The latter function is also called  the inelastic overlap function.
The self-damping of the inelastic channels is responsible for the
transformation of central to a peripheral form of the inelastic channel contribution $h_{inel}(s,b)$ 
with the collision  energy growth  (Fig. 1).
\begin{figure}[h]
\begin{center}
\resizebox{8cm}{!}{\includegraphics*{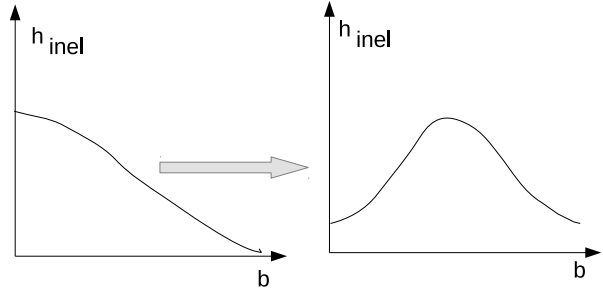}}
\end{center}
\caption[ch1]{\small Transition of the inelastic channel contribution  from  central to a  peripheral profile with an increasing center-of-mass energy.}
\end{figure}
It leads to a dominant role of elastic scattering at $s\to\infty$:
 the cross--section of the inelastic processes rises at $s\to\infty$ with energy as $\ln s$, while
  elastic and total cross--sections behave asymptotically as $\ln^2 s$.
  
  Beyond the new mode threshold energy $s_0$ the two regions in the impact
 parameter space coexist: the central region where self-damping of inelastic channels takes place
 and the peripheral region $b> r(s)$ of the
shadow or absorptive scattering. The function $r(s)$ is determined by the equation $\kappa(s,b=r(s))=0$.
 The collisions at the small impact parameter scattering
are almost elastic ones at the energies ${s}\gg {s_0}$. But at the LHC energies,  
the value of $r(s)$ is small and the elastic amplitude
only slightly exceeds the black disk limit in head-on collisions at $b<r(s)$ at the LHC energies \cite{alkin}.

The head--on colliding particles, at the energies far beyond the LHC ones, 
would provide appearance
of a long-penetrating component in the  EAS
 and such particles would  spend relatively
small part  of
their energy for the production of secondaries. The head-on collisions
would lead therefore to a smaller  number of secondary particle providing
a steeper decrease of the energy spectrum, i.e. it
  will result in the appearance of the knee.  The self-damping leads to suppression of particle
production processes at small impact parameters and
it would be responsible for the slow down of the energy
dependence of the average multiplicity  \cite{avmult}.

The schematic energy dependence of the function $P(s,b=0)$
 is shown in Fig. 2. Various particular models can be utilized for calculation of the function $P(s,b)$ and they would lead to the variations in the particular functional  energy dependence, but the distinct feature related to the all models 
{\it with the reflective scattering mode} is the non--monotonous energy dependence of the function $ P(s,b=0)$
 with its minimum at the 
energy $s_0$.
\begin{figure}[h]
\begin{center}
\resizebox{8cm}{!}{\includegraphics*{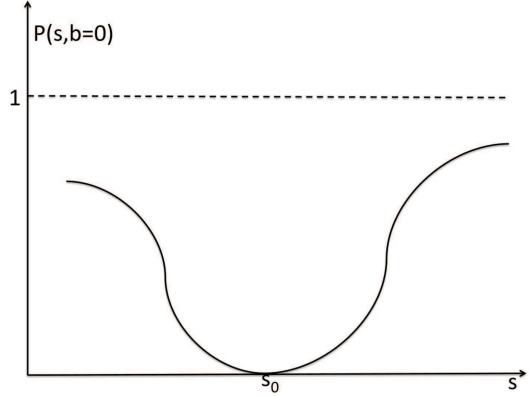}}
\end{center}
\caption[ch1]{\small A schematic energy dependence of the function $P(s,b=0)$ in the typical models with reflective scattering.}
\end{figure} 

As it was noted, the reflective scattering mode emerges
at  the LHC energies as it was predicted in \cite{phl1} and  independently in \cite{degr}, it was confirmed on the base
of the experimental data analysis in  \cite{alkin}, and studies of the EAS
originated from the primary cosmic particles interactions
with atmosphere  provide further evidence for it. 
The main contribution to
the mean multiplicity comes from
the region of $b\sim r(s)$ leading to the events with
 the alignment which also has been observed in EAS \cite{lipari}.  

The detected EAS composition  is correlated with the behavior of the function $P(s,b=0)$.
 It is clear that its larger value
means higher fraction of the elastic (proton)  component. 
Therefore, increase of this component would result  in the enhancement of
the relative fraction
of the protons in  the observed cosmic rays spectrum since just the protons mainly 
constitute  the primary cosmic rays. 
Otherwise, decrease of $P(s,b=0)$ means the
increase of the pion  component in EAS and consequently an increasing
number of muons  observed in form of  the multi-muon events.
The experimental data revealed that relative fraction of protons in cosmic rays shows
non-monotonous energy dependence which is similar to $P(s,b=0)$ dependence (cf. Fig. 3). 
\begin{figure}[thb]
\begin{center}
\includegraphics[width=80mm]{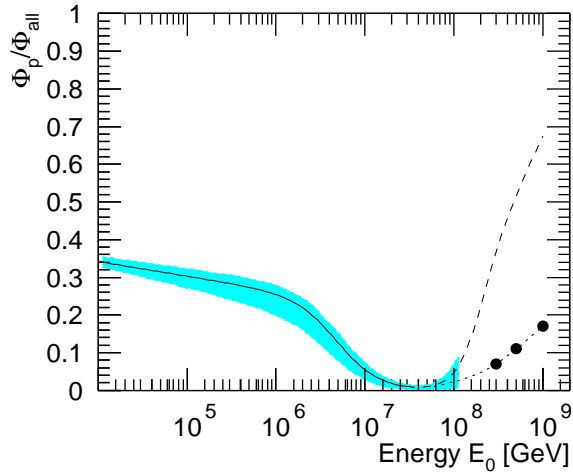}
\end{center}
\small{\caption{Relative fraction of the protons in EAS,
figure is taken from  \cite{hern}; the further related details can be found there.}}
\label{pfrac}
\end{figure}

\section{Conclusion}
We considered  an approach where
the corresponding particle generation mechanism in EAS is  affected by the unitarity effects, and
 the energy region between the knee and the ankle coincides with
 the region where the  reflective scattering mode becomes more and more prominent.  
 
 Other interesting cosmic rays phenomena such as Centauro and anti-Centauro events  have also been observed, 
but those cannot be explained solely by the emergent
reflective scattering. The reflective scattering itself needs to be interpreted at the dynamical level and such interpretation could
provide, in particular, a hint to explanation of the other cosmic peculiarities.

 The distinctive feature of the reflective scattering mode is the peripheral impact parameter distribution of the probability 
of the inelastic production. This can be associated with the production of the hollow fireball in the intermediate state of hadron--hadron interaction.
The projection of this fireball onto the transverse plane looks like a black ring.
The interpretation of this hollow fireball can be borrowed from the papers written about two decades ago \cite{bj,bj1}.
It was supposed  that the interior of this fireball filled by the disoriented chiral condensate which finally irradiates
away. This irradiation is supposed to be coherent, classical with a given isospin in each event, i.e. in one event the isospin can be 
directed along with $\pi^0$ while in other event its direction can be orthogonal leading to production of charged pions. It would explain
appearance of the Centauro and anti-Centauro events.  
 
Thus, the emergent reflective scattering mode could provide a clue for the understanding of the 
peculiarities observed in the EAS studies
of the cosmic radiation. The further studies of the proton scattering in the forward
 region at the LHC will be very helpful for the improved 
 interpretation of the results of the cosmic rays experiments.

\section*{Acknowledgement}
We are grateful to V.A. Petrov for the interesting discussions of the reflective scattering mode.

\end{document}